\documentclass[useAMS,usenatbib]{mn2e}

\usepackage{graphicx}
\usepackage{bm}

\title[Mira Hybrid Wind]{Is Mira a magneto-dusty rotator?}
\author[Anand Thirumalai and Jeremy S. Heyl]{Anand Thirumalai$^{1}$\thanks{E-
mail:anand@phas.ubc.ca (AT); heyl@phas.ubc.ca (JSH)} and Jeremy S. Heyl$^{1}$
\footnotemark[1]\\
$^{1}$University of British Columbia, 6224 Agricultural Road, Vancovuer, British Columbia, 
V6T 1Z1, Canada}

\begin{document}

\date{\today}

\pagerange{\pageref{firstpage}--\pageref{lastpage}} \pubyear{2011}

\maketitle

\label{firstpage}

\def\aj{AJ}                   % Astronomical Journal
\def\apj{ApJ}                 % Astrophysical Journal
\def\apjl{ApJL}                % Astrophysical Journal, Letters
\def\aap{Astron. \&  Astroph.}                % Astronomy and Astrophysics
\def\araa{ARA\&A}             % Annual Review Astronomy and Astrophysics
\def\apjs{ApJ}                % Astrophysical Journal Supplement Series
\def\apss{ApSS}		%Astrophysics and Space Science
\def\mnras{MNRAS}             % Monthly Notices of the RAS
\def\nat{Nature}              % Nature
\def\physrep{Phys.~Rep.}      % Physics Reports
\def\pra{Phys. Rev. A} 			%Physical Review A
\def\pre{Phys. Rev. E} 			%Physical Review E
\def\prb{Phys. Rev. B} 			%Physical Review B
\def\prd{Phys. Rev. D} 			%Physical Review D
\def\pasp{Pubs. Astron. Soc. Pac.}
\def\solphys{Sol. Phys.}

\begin{abstract}
We investigate the possibility that a magnetic field may be present in the star $o-$Ceti (hereafter, Mira) and that the field plays a role in the star's mass loss. The model presented here is an application of an earlier derived theory that has been successfully employed for intermediate and high-mass evolved stars, and is now extended to the low-mass end. The modelling shows that it is possible to obtain a hybrid magnetohydrodynamic-dust-driven wind scenario for Mira, in which the role of a magnetic field in the equatorial plane of the star is dynamically important for producing a stellar wind. The wind velocity and the temperatures obtained from the model appear consistent with findings elsewhere.
\end{abstract}


\section{Introduction}\label{sec:intro}
Mira (omicron-Ceti) is a relatively close evolved star that is in its asymptotic giant branch (AGB) infancy. It is the template upon which the class of stars called \emph{Mira variables} are based, which are characterised most strikingly, by large-amplitude long period variability. Mira varies with a period of about $332$ days \citep[e.g.][]{Lopez1997}. Observations of emission due to vibrational-rotational transitions in the CO molecule have not only revealed that the envelope of Mira is mildly asymmetric, possibly due to a bipolar outflow, perhaps due to an equatorial magnetic field \citep[see for e.g.][]{Planesas1990a,Planesas1990b, Lopez1997} but also CO transition line observations have also enabled a variety of different estimates of the wind velocity ranging from $4.8$ km/s \citep[see][]{Young1995} to $2.5$ km/s for the envelope's expansion at intermediate distances of $100-1000 R_0$ \citep[see][]{Ryde2000,Ryde2001}. Elsewhere, there is evidence for double winds \citep[see][]{Knapp1998} with a fast wind of about $6.7$ km/s, and a slow component of about $2.5$ km/s. Meanwhile, observations in the infrared at around $11 \mu$m, have revealed that the inner radius for the dust around Mira is at around $3 R_0$ \citep[see for e.g.][]{Bester1991,Lopez1997}, where $R_0$ is the photospheric radius. These observations have also been modelled with success \citep[see][]{Lopez1997} using an additional dust shell at $12 R_0$, as well as with dust clumps with photospheric spots. 

SiO-maser observations of the molecular shell close to the photosphere of Mira have revealed that the star may be rotating with a period of about $89 \times \sin(i)$ years \citep[see][]{Cotton2006}, where $i$ is the angle between the line of sight and the rotation axis. SiO maser polarisation studies on the other hand, indicate that Mira may harbour a predominantly radial magnetic field in its atmosphere \citep[e.g.][]{Cotton2004,Cotton2006}. Elsewhere, recent observations of masers have been instrumental in establishing the dynamic role that magnetic fields play in the atmospheres of evolved stars \citep[see for e.g.][]{Amiri2012,Vlemmings2011,Herpin2006} and theoretical efforts \citep[see][]{Busso2007} have delineated their role in transport of material in AGB stellar interiors with surface fields of $\stackrel{_<}{_\sim} 20$~G. 

Imaging of Mira in the near infrared (IR) and optical wavelengths has provided tools for measuring the diameter of the star \citep[see for e.g.][]{Haniff1995, van_Leeuwen1997}.  \cite{Perrin2004} find that by varying the opacity of the molecular layer just ahead of the photosphere, they were able to account for the apparent changes in the diameter of Mira, arriving at an estimate of $\approx 350 R_{\odot}$, over the entire variability cycle of Mira. 

From the brief discussion of the literature presented above, it appears that the role that magnetic fields may play in shaping the outflow and influencing certain dynamic features of the envelope cannot be ignored. In the current study we present a rudimentary model integrating the effects of rotation, an equatorial magnetic field and the usual dust-driving picture into one cohesive scenario for Mira's outflow. The work presented here represents an extension of our earlier theory \citep[see][]{Thirumalai2010, TH2012} to the low-mass end of AGB stars; viz., Mira which is about $1.5 M_{\odot}$ \citep[e.g.][]{Martin2007}. Now that stellar rotation may have been detected in Mira, the aim of the current work is to raise the question, \emph{how important is the magnetic field in the stellar outflow and can Mira be a magneto-dusty rotator, given the current observations?}

\section{The Hybrid-Wind Model}\label{sec:model}

We confine our attention to an axisymmetric model in the equatorial plane of the star. Therein, the velocity fields of the gas and the dust as well as the magnetic field are assumed to have radial and azimuthal components. However, the azimuthal components are functions of purely the radial distance from the star; this is the central assumption behind the canonical Weber-Davis  \citep[see][hereafter WD]{WD67} model for our Sun. There are two fluids; the gas which carries the magnetic field and secondly the dust, which moves through the gas, dragging the gas with it. The two fluids are coupled by a drag term in the usual way. The dust-to-gas ratio is kept small around $\sim 1/355$ \citep[e.g.][]{Danchi1994} and the individual dust grains are assumed to be spherical in shape. The gas is assumed to have a polytropic equation of state, with a polytropic exponent $\gamma > 1$ \citep[see][]{Thirumalai2010, TH2012}, where a value of unity represents the isothermal limit. We assume ideal MHD where there is no Lorentz force in the fluid. The input parameters for model of Mira's wind are listed in Table~1. A steady-state description of the gas velocity in the hybrid wind in the equatorial plane of Mira can be written as \citep[see][]{Thirumalai2010},
\begin{eqnarray}
\frac{dw}{dx}=\frac{w}{x}\frac{N(w,x)}{D(w,x)}~,
\label{eq:1}
\end{eqnarray}
where, $w=u/u_A$ is the gas speed normalised using the Alfv\'{e}n
speed and $x=r/r_A$, is the radial distance expressed in units of the
Alfv\'{e}n radius. Hereafter, the subscript `$A$' refers to values of the different variables at the 
Alfv\'{e}n radius. The quantities $N(w,x)$ and $D(w,x)$ are the
numerator and denominator respectively and are given by,
\begin{eqnarray}
N(w,x)= \left(2 \gamma S_T (wx^2)^{1-\gamma} - \frac{S_G}{x}(1-\Gamma_d \cdot \Theta(x-
x_d))\right) \nonumber\\
\times (wx^2-1)^3 +~ S_{\Omega}x^2(w-1)\left(1-3wx^2 + (wx^2+1)w \right)
\label{eq:2} 
\end{eqnarray}
and
\begin{eqnarray}
D(w,x)= \left(w^2-\gamma S_T (wx^2)^{1-\gamma}\right)(wx^2-1)^3-
S_{\Omega}x^2 \times \nonumber\\
(wx^2)^2\left(\frac{1}{x^2}-1\right)^2.
\label{eq:3}
\end{eqnarray}
In the above equations, the parameters $S_T=\frac{2kT_A}{m_p u_A^2}$,
$S_G=\frac{GM_*}{r_A u_A^2}$ and $S_\Omega=\frac{\Omega^2
  r_A^2}{u_A^2}$ along with $\gamma$ uniquely determine the locations
of the critical points, and hence the morphology of the family of
solutions of Eq.~(\ref{eq:1}). Here $T_A$ is the gas temperature at the Alfv\'{e}n radius, $k$ is 
the Boltzmann constant while $m_p$ is the mass of a proton. The presence of the Heaviside function in
Eq.~(\ref{eq:2}) represents the formation of dust at the location
$x=x_d$. The dust velocity profile is then given by \citep
[see][]{Thirumalai2010},
\begin{equation}
v(r)=u(r)+\left(\frac{\sqrt{a_{th}^4+4\left(\frac{\Gamma_d GM_{*}}{\pi a^2 n_d r^2}\right)^2}-a_
{th}^2}{2}\right)^{1/2},
\label{eq:4}
\end{equation}
where $a_{th}$ is the thermal speed given by $a_{th}=\sqrt{2kT/\mu m_u}$ and $\mu m_u$ is 
the mean molecular mass of the gas and $n_d$ is the dust grain number density, which is 
assumed to be given by, $n_d m_d / \rho \approx \langle\delta\rangle$, with $\langle\delta
\rangle$, the average dust-to-gas ratio in the wind. Eq.~(\ref{eq:1}) is solved numerically on a computer \citep[see][]{Thirumalai2010,TH2012}. In the following section the results are presented alongside a discussion.
\begin{table}
 \centering
\label{tab:Table1}
%\begin{minipage}{140mm}
  \caption{Various parameters for modelling Mira. The variable parameters are listed as such.}
  \begin{tabular}{@{}lcl@{}}
  \hline
   Parameter     & Symbol        &  Value / Comment \\
 \hline
Mass & $M_*$ & $\sim 1.5\mathrm{M}_{\odot}$ \\
Radius & $R_0$ & $\sim 464\mathrm{R}_{\odot}$ \\
Mass loss rate & $\dot{M}$ & $\sim 3 \times 10^{-7} \mathrm{M}_{\odot}$/yr \\
Surface magnetic & & \\ 
field strength & $B_0$ & variable \\ 
Bulk surface gas & & \\ 
velocity (radial) & $u_0$ & variable\\
Surface temperature (effective) & $T_0$ & $\sim 2500$K \\
Stellar rotation rate & $\Omega$ & $2\pi/\left [ 89 \times \sin(i)
  ~\mathrm{yr} \right]$ \\
Rotation axis angle & $i$ & variable \\
Surface escape velocity & $v_\mathrm{esc,0}$ & $3.52 \times 10^6$ cm/s \\
Polytropic exponent & $\gamma$ & $> 1$ \\
Alfv\'en radius & $r_A$ & variable \\
Alfv\'en speed & $u_A$ & variable \\
Dust Parameter & $\Gamma_d$ & variable \\
Dust grain radius & a & spherical grains \\
Dust grain mass & $m_d$ & $\sim 4/3 \pi a^3 \rho_d$ \\
\hline
\end{tabular}
%\end{minipage}
\end{table}
 
\section{Results and Discussion}\label{sec:results}

Along with the basic ingredients of the model listed in Table~1, the set of parameters $\{ B_0, u_0, u_a, r_a, \gamma, \Gamma_d, i \}$ are varied until a critical solution to Eq.~(\ref{eq:1}) is obtained that satisfies the following criteria.

\begin{enumerate}
	\item[1.]{The solution passes through all three critical points; the sonic point, the radial Alfv\'{e}n point and the fast point}
	\item[2.]{The solution is continuous through the radial Alfven point.}
	\item[3.]{The velocity profile starts at the base of the wind sub-sonic and attains a super-Alfv\'{e}nic terminal velocity at large distances}
	\item[4.]{The temperature range in the dust condensation region (within a few stellar radii from the photosphere) is consistent with observations.}
	\item[5.]{The gas terminal velocity is consistent with observations.}
\end{enumerate} 

\begin{figure}
\begin{center}
\includegraphics[width=3.5in, scale=0.9]{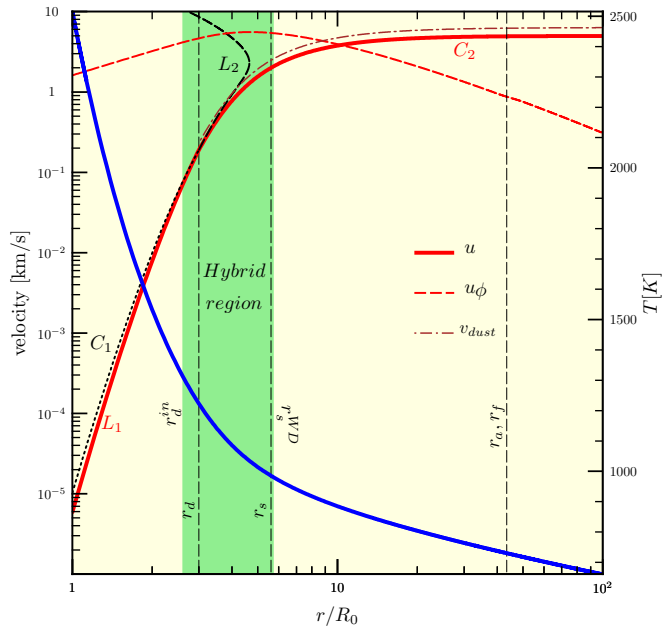}
\end{center}
\caption{Hybrid wind solution is shown for Mira with parameters $u_A \approx 0.14 v_{esc,
0}$, 
$r_A \approx 43.47 R_0$ and for $\Gamma_d \approx 0.06$ and remaining parameters as given in
Tables 1 and 2.  The red solid line ($L_1+C_2$) traces the hybrid MHD-dust-driven wind solution for Mira. The decreasing blue solid line traces the temperature and should be interpreted using the right hand y-axis.} 
\label{fig:figure1}
\end{figure}

This optimisation procedure carried out in tandem with integrating the differential equation in Eq.~(\ref{eq:1}) results in a picture of a hybrid MHD-dust-driven wind for Mira as shown in Figure~ \ref{fig:figure1}. The critical wind solution is shown as the solid red line, which comprises of two parts, $L_1$ and $C_2$. These two lines intersect at the dust formation radius $r_d$, which in the current model is located at $3 R_0$. In the region $r \geq r_d$ Eq.~(\ref{eq:1}) is integrated with the presence of the dust parameter $\Gamma_d$, which represents dust grain drag acting on the gas; it is this part of the solution that is labelled $C_2$. On the other hand, inside $r \leq r_d$ there is no dust and as a result Eq.~(\ref{eq:1}) is integrated in this latter region without the dust parameter, as a pure WD solution. This part of the solution is labelled as $L_1$. Thus in the region $R_0 \leq r \leq r_d$, a pure WD mechanism is responsible for transport of stellar material. Together, $L_1 + C_2$ forms the combined hybrid MHD-dust-driven wind. 
The radial Alfv\'{e}n point and the fast point lie nearly coincident upon one another; shown by the black dashed vertical line at $r \approx 43 R_0$. Theoretically speaking, if we assume that dust condensation occurs at the photosphere, then we would obtain the combined solution $C_1 + C_2$ by integrating Eq.~(\ref{eq:1}) with the dust parameter, over the entire domain $R_0 \leq r \leq 100 R_0$. However, in the atmosphere of Mira, dust formation occurs at the dust condensation radius $r=r_d$ and not at the photosphere. Therefore, inside the dust radius, $r \leq r_d$, the wind starts off at the photosphere at some velocity $u_0$, and proceeds outwards along the trajectory $L_1$, and after dust condensation at $r=r_d$, the solution then switches to proceed outward along $C_2$, rather than continue along the unphysical solution $L_2$. 

Thus, it is seen explicitly, that without the onset of dust formation at $r=r_d$, there would not be any efflux from Mira, since there is only one solution ($C_2$) that passes through all three critical points, and emerges super-Alfv\'{e}nic at large distances. As can be seen, the gas terminal velocity is about 5~km/s in the equatorial plane, which is in reasonably good agreement with estimates of the wind velocity of Mira \citep[e.g.][]{Young1995,Martin2007}. 

The dust velocity profile is also shown in Figure~\ref{fig:figure1} as the brown dot-dashed line. This is computed according to Eq.~(\ref{eq:4}), once the gas velocity profile is known. The dust velocity is slightly larger than the gas velocity, as expected, in a dust-driven wind. The red dashed line represent the azimuthal velocity profile of the gas. This profile is typical for a magneto-centrifugal wind. The blue solid line shows the temperature profile in the atmosphere of Mira and should be interpreted using the right hand side axis. The photospheric temperature is about $2500$ K for the model shown. Finally, the green shaded area shows the so called \emph{hybrid region}. This region is bounded to the left by $r=r_d^{in} \approx 2.6 R_0$, this is the lower limit for the inner dust radius as given by \cite{Danchi1994}. To the right, the hybrid region is bounded by the sonic point of the pure WD magneto-centrifugal model. The sonic point of the hybrid model lies at $r \approx 5.61 R_0$; just inside the hybrid region. As seen in our earlier study \citep[see][]{Thirumalai2010}, one of the ways in which a hybrid wind is possible is if the sonic point of the hybrid model lies within the sonic point of the pure WD model; i.e., $r_s < r_s^{WD}$. Moreover, the dust formation radius must also then lie inside $r \leq r_s$. The temperature in the hybrid region can be inferred from the temperature profile shown, to be about $1000 \stackrel{_<}{_\sim} T \stackrel{_<}{_\sim} 1300 $ K, which is well within the observed range of temperatures at this distance from the star \citep[e.g.][]{Lopez1997, Bester1991, Danchi1994, Perrin2004}. 

Finally the optimised values for the different variable parameters in the model are given in Table 2. Notice that the surface radial magnetic field at the photosphere is obtained to be about $4$ G, which is within the range of field strengths estimated by \cite{Herpin2006} and \cite{Busso2007} for AGB stars. 

\begin{figure}
\begin{center}
\includegraphics[width=3.5in, scale=0.9]{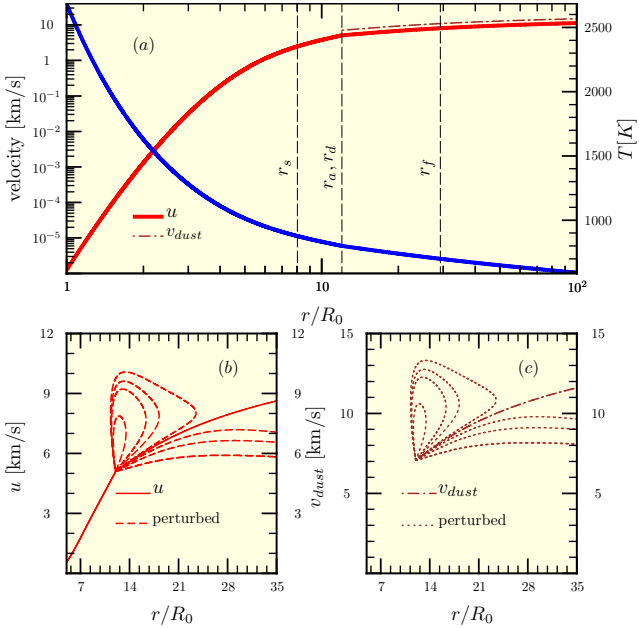}
\end{center}
\caption{(a) Hybrid wind solution is shown for a scenario with a hot spot at the photosphere of Mira. This model has parameters $u_A \approx 0.15v_{esc,
0}$, 
$r_A \approx 29.21 R_0$ and $\Gamma_d \approx 0.1$ and remaining parameters as given in
Table 1. The surface magnetic field was found to be $B_0 \approx 1.15$G. The decreasing blue solid line traces the temperature and should be interpreted using the right hand y-axis. (b) Gas velocities obtained by perturbing the solution in the vicinity of $r_A$ and (c) the corresponding dust velocities.} 
\label{fig:figure2}
\end{figure}

Presently, we turn our attention to the question of hot spots on the photosphere of Mira, and the related question of the influence of the spot on the stellar wind, ahead of the spot in the atmosphere. Figure~\ref{fig:figure2}(a) shows a hybid MHD-dust-driven wind model where the photospheric temperature is increased to about $2700$ K. However, in this second model instead of formulating a hybrid wind model as before, with dust formation occurring within the sonic point, we chose a different scenario. Our motivation here was to investigate the possibility regarding dust formation at around $12 R_0$, at which distance \cite{Lopez1997} model a second dust shell. 
Within the framework of the current theory, we found that the only way to have a viable wind scenario with dust formation at this distance, was by collocating the dust formation radius with the radial Alfv\'{e}n point. However, in this case, as can be seen from Figure~\ref{fig:figure2}, the solution is not smooth at the radial Alfv\'{e}n point. In this scenario, the solution need not be smooth at $r=r_A$, as this also happens to be the dust formation radius, hence a kink in the solution is allowed. It is possible that the dust formation can occur outside the fast point, by fine tuning the parameters of the model. This would place the dust formation radius in the vicinity of about $20-30 R_0$. While we do not show such a scenario in Figure~\ref{fig:figure2}, it is to be acknowledged that it may be possible within the framework of the hybrid MHD-dust-driven wind theory. Our objective here was to formulate a model with dust formation occurring at a distance of about $12 R_0$, to be consistent with other models for dusty shells or clumps \citep[e.g.][]{Lopez1997}, while obtaining temperatures in agreement with observations.

In Figure~\ref{fig:figure1} it can be seen that the temperature at the left bound of the hybrid region ($r=r_d^{in}=2.6 R_0$) is about $\sim 1300$ K. This same temperature is calculated to be at a distance of about $2.9 R_0$ in Figure~\ref{fig:figure2}. Thus, ahead of a hot spot in the atmosphere, the viable dust formation region moves further out from the photosphere, as would be expected. In Figure~\ref{fig:figure2} the temperature at the sonic point is seen to be about $\sim 900$ K. Thus the range of temperature in this model is similar to that for the hybrid region in Figure~\ref{fig:figure1}, therefore it is still possible to formulate a hybrid wind scenario with dust formation occurring within the sonic point, as was shown in Figure~\ref{fig:figure1}. However, as mentioned earlier, in the current model our goal is to locate a second dust shell outside the sonic point at around $12R_0$ \citep[see][]{Lopez1997} and not within it. 

In this scenario, with a short-lived hot spot on the photosphere, the temperature in the gas is about $800$ K at the radial Alfv\'{e}n point ($r=r_A$). Thus, the observed temperature for the dust of about $600$ K at this distance is quite feasible \citep[e.g.][]{Lopez1997}. Overall, for the lifetime of the hot spot, it is possible to sustain a hybrid wind with dust forming at $12 R_0$. A pure WD mechanism transports stellar material from the photosphere, through the sonic point and out to the radial Alfv\'{e}n point. After dust formation at this location, the MHD-dust-driven wind then negotiates its way through the remaining critical point; the fast point and leaves the star super-Alfv\'{e}nic. Comparing Figure~\ref{fig:figure2}(a) with Figure~\ref{fig:figure1} reveals that the hybrid model is quite sensitive to changes in the photospheric temperature and the magnetic field. For the model shown in Figure~\ref{fig:figure2}, the surface radial magnetic field strength was found to be $\approx 1.15$ G. The dust parameter ($\Gamma_d$) was found to be about $0.1$; a little higher than in the previous model (c.f. Table~2). With these parameters, the terminal wind velocity is also concomitantly higher, at around $11$ km/s in the equatorial plane. In comparison to the time for one stellar rotation, since the hot-spot at the photosphere would be short-lived, therefore the hot spot model described above would be valid for the duration of the spot. 

In this regard, there are some concerns that are inherited from collocating the dust formation and radial Alfv\'{e}n points. In the main, dust formation and growth is a stochastic process that is not completely understood. Particularly, for the wind ahead of a short-lived hot spot in the atmosphere, the physical processes may be quite dynamic over time-scales comparable to the spot lifetime. As a result small changes in the wind velocity in the vicinity of the radial Alfv\'{e}n point can result in a drastically different wind velocity profile and indeed the wind may not be able to navigate through the fast point. This is shown in Figure~\ref{fig:figure2}(b) and (c). We have perturbed the gas velocity by a small amount ($\stackrel{_<}{_\sim} 0.01\%$) ahead of the radial Alfv\'{e}n point and then integrated the perturbed solution outwards. Figure~\ref{fig:figure2}(b) shows the unperturbed solution as the red solid line and the perturbed wind solutions as the dashed lines. We see that this perturbation of the gas velocity results in the wind becoming either a failed wind, where it does not pass through the fast point (the dashed lines below the solid red line) or the wind becomes an unphysical double valued wind (dashed lines above the the red solid line that form loops). In either case, this suggests that minor perturbations of the wind velocity around the radial Alfv\'{e}n point will drastically change the nature of the outflow. Figure~\ref{fig:figure2}(c) shows the corresponding dust velocity profile for both the unperturbed and the perturbed solutions. Should the dust follow any of the perturbed trajectories then it will not leave the star and can lead to instabilities in the flow. Therefore, while it is possible that in the atmosphere of Mira, conditions ahead of a mild hot spot may result in a hybrid wind with dust formation at around $12 R_0$, these conditions would be limited by the life-time of the spot, and as such cannot be classified as steady-state, even when compared to the stellar rotation time. Such a localised phenomenon ahead of a hot spot may be able to account for clumpiness of the dust distribution and even spatially distinct dust shells around Mira \citep[e.g.][]{Lopez1997}. We shall end this discussion by cautioning the reader that fully dynamic 2- or 3-D modelling would be needed for establishing the importance of such dynamic short time-scale phenomena, which cannot be entirely gleaned from the simplistic theory presented here.
\begin{table}
 \centering
\label{tab:Table2}
%\begin{minipage}{140mm}
  \caption{Optimised values of the variable parameters for the star Mira.}
  \begin{tabular}{@{}ll@{}}
  \hline
  \hline
   Parameter     &  Value  \\
 \hline
 \hline
$B_0$ & $\approx 4.11$ G\\
$u_0$ & $\approx 5.89 \times 10^{-6}$ km/s \\ 
$u_A$ & $\approx 4.93$ km/s \\
$r_A$  & $\approx 43.47 R_0$ \\
$\gamma$ & $\approx 1.06$ \\
$\Gamma_d$ & $\approx 0.06$ \\
$i$ & $\approx 26.60^{\circ} \approx 0.46 $ rad \\ 
\hline
\hline
\end{tabular}
%\end{minipage}
\end{table}
The conclusions of the current work are summarised in the following section.

\section{\label{sec:conc}Conclusion}

In this letter we have modelled the stellar wind of Mira as a hybrid MHD-dust-driven wind. The study was motivated largely by hints of the discovery of a magnetic field in the star in addition to a possible detection of stellar rotation. 

Since the surface magnetic field strength as well as the rotation rate are not exactly known, these along with a few other physical parameters such as the dust parameter ($\Gamma_d$) were treated as variable parameters that were fine-tuned, in order to arrive at a suitable stellar wind model for Mira. We obtained a stellar wind with a terminal velocity of about 5~km/s in the equatorial plane. This value for the wind velocity is in agreement with current estimates. The stellar effective temperature was taken to be about 2500~K and we obtained a surface magnetic field of about $\approx 4$~G, in good agreement with current estimates. We also arrived at an estimate for a \emph{hybrid region} in the inner atmosphere of Mira. This region depicts the range of distances at which dust condensation can occur in order to produce a hybrid wind. The temperature profile obtained for this hybrid region is consistent with models of the atmosphere of Mira.

For the purpose of modelling dust shells around Mira at several stellar radii, we employed a model of a hot-spot that alters the wind dynamics ahead of the spot in the atmosphere. With this we were able to show that it may be possible to facilitate dust formation at around $12 R_0$ by collocating the dust formation and radial Alfv\'{e}n points, resulting in a hybrid wind. However, because such a spot may be short-lived the dust and gas velocities in the vicinity of the radial Alfv\'{e}n point are likely to change over short timescales, resulting in failed wind solutions and possibly dynamic instabilities.  This may form form localised regions of dusty clumps, or even distinct shells with slightly enhanced dust density. 

It is however to be remembered, that the current model is an idealised steady-state description of a more complicated and dynamic problem. As such, the results obtained here only convey the possibility, that Mira may be a magneto-dusty rotator. Dynamic multi-D modelling of its wind is required for obtaining ultimately, a truer description of the nature of its efflux and of its atmosphere.

\section*{Acknowledgements}
This research was supported by funding from NSERC.  The calculations
were performed on computing infrastructure purchased with funds from
the Canadian Foundation for Innovation and the British Columbia
Knowledge Development Fund. The authors are also grateful to the referee for useful discussions for improving the paper.

\bibliographystyle{mn2e}
\bibliography{mira}

\begin{thebibliography}{}

\bibitem[\protect\citeauthoryear{{Amiri}, {Vlemmings}, {Kemball} \& {van
  Langevelde}}{{Amiri} et~al.}{2012}]{Amiri2012}
{Amiri} N.,  {Vlemmings} W.~H.~T.,  {Kemball} A.~J.,    {van Langevelde} H.~J.,
   2012, \aap, 538, A136

\bibitem[\protect\citeauthoryear{{Bester}, {Danchi}, {Degiacomi}, {Townes} \&
  {Geballe}}{{Bester} et~al.}{1991}]{Bester1991}
{Bester} M.,  {Danchi} W.~C.,  {Degiacomi} C.~G.,  {Townes} C.~H.,    {Geballe}
  T.~R.,  1991, \apjl, 367, L27

\bibitem[\protect\citeauthoryear{{Busso}, {Wasserburg}, {Nollett} \&
  {Calandra}}{{Busso} et~al.}{2007}]{Busso2007}
{Busso} M.,  {Wasserburg} G.~J.,  {Nollett} K.~M.,    {Calandra} A.,  2007,
  \apj, 671, 802

\bibitem[\protect\citeauthoryear{{Cotton}, {Mennesson}, {Diamond}, {Perrin},
  {Coud{\'e} du Foresto}, {Chagnon}, {van Langevelde}, {Ridgway}, {Waters},
  {Vlemmings}, {Morel}, {Traub}, {Carleton} \& {Lacasse}}{{Cotton}
  et~al.}{2004}]{Cotton2004}
{Cotton} W.~D.,  {Mennesson} B.,  {Diamond} P.~J.,  {Perrin} G.,  {Coud{\'e} du
  Foresto} V.,  {Chagnon} G.,  {van Langevelde} H.~J.,  {Ridgway} S.,  {Waters}
  R.,  {Vlemmings} W.,  {Morel} S.,  {Traub} W.,  {Carleton} N.,    {Lacasse}
  M.,  2004, \aap, 414, 275

\bibitem[\protect\citeauthoryear{{Cotton}, {Vlemmings}, {Mennesson}, {Perrin},
  {Coud{\'e} du Foresto}, {Chagnon}, {Diamond}, {van Langevelde}, {Bakker},
  {Ridgway}, {McAllister}, {Traub} \& {Ragland}}{{Cotton}
  et~al.}{2006}]{Cotton2006}
{Cotton} W.~D.,  {Vlemmings} W.,  {Mennesson} B.,  {Perrin} G.,  {Coud{\'e} du
  Foresto} V.,  {Chagnon} G.,  {Diamond} P.~J.,  {van Langevelde} H.~J.,
  {Bakker} E.,  {Ridgway} S.,  {McAllister} H.,  {Traub} W.,    {Ragland} S.,
  2006, \aap, 456, 339

\bibitem[\protect\citeauthoryear{{Danchi}, {Bester}, {Degiacomi}, {Greenhill}
  \& {Townes}}{{Danchi} et~al.}{1994}]{Danchi1994}
{Danchi} W.~C.,  {Bester} M.,  {Degiacomi} C.~G.,  {Greenhill} L.~J.,
  {Townes} C.~H.,  1994, \aj, 107, 1469

\bibitem[\protect\citeauthoryear{{Haniff}, {Scholz} \& {Tuthill}}{{Haniff}
  et~al.}{1995}]{Haniff1995}
{Haniff} C.~A.,  {Scholz} M.,    {Tuthill} P.~G.,  1995, \mnras, 276, 640

\bibitem[\protect\citeauthoryear{{Herpin}, {Baudry}, {Thum}, {Morris} \&
  {Wiesemeyer}}{{Herpin} et~al.}{2006}]{Herpin2006}
{Herpin} F.,  {Baudry} A.,  {Thum} C.,  {Morris} D.,    {Wiesemeyer} H.,  2006,
  \aap, 450, 667

\bibitem[\protect\citeauthoryear{{Knapp}, {Young}, {Lee} \& {Jorissen}}{{Knapp}
  et~al.}{1998}]{Knapp1998}
{Knapp} G.~R.,  {Young} K.,  {Lee} E.,    {Jorissen} A.,  1998, \apjs, 117, 209

\bibitem[\protect\citeauthoryear{{Lopez}, {Danchi}, {Bester}, {Hale}, {Lipman},
  {Monnier}, {Tuthill}, {Townes}, {Degiacomi}, {Geballe}, {Greenhill},
  {Cruzalebes}, {Lefevre}, {Mekarina}, {Mattei}, {Nishimoto} \&
  {Kervin}}{{Lopez} et~al.}{1997}]{Lopez1997}
{Lopez} B.,  {Danchi} W.~C.,  {Bester} M.,  {Hale} D.~D.~S.,  {Lipman} E.~A.,
  {Monnier} J.~D.,  {Tuthill} P.~G.,  {Townes} C.~H.,  {Degiacomi} C.~G.,
  {Geballe} T.~R.,  {Greenhill} L.~J.,  {Cruzalebes} P.,  {Lefevre} J.,
  {Mekarina} D.,  {Mattei} J.~A.,  {Nishimoto} D.,    {Kervin} P.~W.,  1997,
  \apj, 488, 807

\bibitem[\protect\citeauthoryear{{Martin}, {Seibert}, {Neill}, {Schiminovich},
  {Forster}, {Rich}, {Welsh}, {Madore}, {Wheatley}, {Morrissey} \&
  {Barlow}}{{Martin} et~al.}{2007}]{Martin2007}
{Martin} D.~C.,  {Seibert} M.,  {Neill} J.~D.,  {Schiminovich} D.,  {Forster}
  K.,  {Rich} R.~M.,  {Welsh} B.~Y.,  {Madore} B.~F.,  {Wheatley} J.~M.,
  {Morrissey} P.,    {Barlow} T.~A.,  2007, \nat, 448, 780

\bibitem[\protect\citeauthoryear{{Perrin}, {Ridgway}, {Mennesson}, {Cotton},
  {Woillez}, {Verhoelst}, {Schuller}, {Coud{\'e} du Foresto}, {Traub},
  {Millan-Gabet} \& {Lacasse}}{{Perrin} et~al.}{2004}]{Perrin2004}
{Perrin} G.,  {Ridgway} S.~T.,  {Mennesson} B.,  {Cotton} W.~D.,  {Woillez} J.,
   {Verhoelst} T.,  {Schuller} P.,  {Coud{\'e} du Foresto} V.,  {Traub} W.~A.,
  {Millan-Gabet} R.,    {Lacasse} M.~G.,  2004, \aap, 426, 279

\bibitem[\protect\citeauthoryear{{Planesas}, {Bachiller}, {Martin-Pintado} \&
  {Bujarrabal}}{{Planesas} et~al.}{1990a}]{Planesas1990a}
{Planesas} P.,  {Bachiller} R.,  {Martin-Pintado} J.,    {Bujarrabal} V.,
  1990, \apj, 351, 263

\bibitem[\protect\citeauthoryear{{Planesas}, {Kenney} \&
  {Bachiller}}{{Planesas} et~al.}{1990b}]{Planesas1990b}
{Planesas} P.,  {Kenney} J.~D.~P.,    {Bachiller} R.,  1990, \apjl, 364, L9

\bibitem[\protect\citeauthoryear{{Ryde}, {Gustafsson}, {Eriksson} \&
  {Hinkle}}{{Ryde} et~al.}{2000}]{Ryde2000}
{Ryde} N.,  {Gustafsson} B.,  {Eriksson} K.,    {Hinkle} K.~H.,  2000, \apj,
  545, 945

\bibitem[\protect\citeauthoryear{{Ryde} \& {Sch{\"o}ier}}{{Ryde} \&
  {Sch{\"o}ier}}{2001}]{Ryde2001}
{Ryde} N.,  {Sch{\"o}ier} F.~L.,  2001, \apj, 547, 384

\bibitem[\protect\citeauthoryear{{Thirumalai} \& {Heyl}}{{Thirumalai} \&
  {Heyl}}{2010}]{Thirumalai2010}
{Thirumalai} A.,  {Heyl} J.~S.,  2010, \mnras, 409, 1669

\bibitem[\protect\citeauthoryear{{Thirumalai} \& {Heyl}}{{Thirumalai} \&
  {Heyl}}{2012}]{TH2012}
{Thirumalai} A.,  {Heyl} J.~S.,  2012, Monthly Notices of the Royal
  Astronomical Society, 422, 1272

\bibitem[\protect\citeauthoryear{{van Leeuwen}, {Feast}, {Whitelock} \&
  {Yudin}}{{van Leeuwen} et~al.}{1997}]{van_Leeuwen1997}
{van Leeuwen} F.,  {Feast} M.~W.,  {Whitelock} P.~A.,    {Yudin} B.,  1997,
  \mnras, 287, 955

\bibitem[\protect\citeauthoryear{{Vlemmings}, {Humphreys} \&
  {Franco-Hern{\'a}ndez}}{{Vlemmings} et~al.}{2011}]{Vlemmings2011}
{Vlemmings} W.~H.~T.,  {Humphreys} E.~M.~L.,    {Franco-Hern{\'a}ndez} R.,
  2011, \apj, 728, 149

\bibitem[\protect\citeauthoryear{{Weber} \& {Davis} Jr.}{{Weber} \&
  {Davis}}{1967}]{WD67}
{Weber} E.~J.,  {Davis} Jr. L.,  1967, \apj, 148, 217

\bibitem[\protect\citeauthoryear{{Young}}{{Young}}{1995}]{Young1995}
{Young} K.,  1995, \apj, 445, 872

\end{thebibliography}
\label{lastpage}
\end{document}